\documentclass[aps,twocolumn]{revtex4-2}
\usepackage{amsmath}
\usepackage{amssymb}
\usepackage{graphicx}
\usepackage{array}
\usepackage{dcolumn}
\usepackage{subfigure}
\usepackage{color}
\usepackage{float}
\usepackage{xr-hyper}
\usepackage[hidelinks=true]{hyperref}
\hypersetup{
  colorlinks   = true, %Colours links instead of ugly boxes
  urlcolor     = blue, %Colour for external hyperlinks
  linkcolor    = blue, %Colour of internal links
  citecolor   = red %Colour of citations
}

\usepackage{epsfig}
\usepackage{epstopdf}
\usepackage{amsmath}
\usepackage{amsfonts}
\usepackage{amssymb}
\usepackage{hyperref}
\usepackage{bm}
\usepackage{makecell}
\usepackage{rotating}
\usepackage{hyperref}
\usepackage{multirow}
\usepackage{graphicx}
\usepackage{array} 
\usepackage{tabularx} % make sure this is in your preamble
\newcolumntype{Y}{>{\centering\arraybackslash}X}

\usepackage{chemformula} 
\usepackage[T1]{fontenc} 
\usepackage[percent]{overpic}
\usepackage{academicons} 
\usepackage{xcolor}
\usepackage{ragged2e}

\usepackage{graphicx}% Include figure files
\usepackage{dcolumn}% Align table columns on the decimal point
\usepackage{bm}% bold math
\usepackage{color}
\usepackage{comment}
\usepackage[hidelinks=true]{hyperref}
\usepackage{tikz,xcolor,hyperref}
\DeclareMathAlphabet\mathbfcal{OMS}{cmsy}{b}{n}

\usepackage{physics}

\definecolor{lime}{HTML}{A6CE39}
\DeclareRobustCommand{\orcidicon}{%
	\begin{tikzpicture}
	\draw[lime, fill=lime] (0,0)
	circle [radius=0.16]
	node[white] {{\fontfamily{qag}\selectfont \tiny ID}};
	\draw[white, fill=white] (-0.0625,0.095)
	circle [radius=0.007];
	\end{tikzpicture}
	\hspace{-2mm}
}

\foreach \x in {A, ..., Z}{%
	\expandafter\xdef\csname orcid\x\endcsname{\noexpand\href{https://orcid.org/\csname orcidauthor\x\endcsname}{\noexpand\orcidicon}}
}

\begin{document}
\title{Design of altermagnetism in oxide superlattices exploiting interface effects and quantum confinement}
% \title{Altermagnetism at the interface of electrostatically doped  perovskite superlattices}
% \title{Interface- and confinement-induced altermagnetism in perovskite superlattices}
 
\author{Subhadeep Bandyopadhyay\orcidA}
\email{subha.7491@gmail.com}
\affiliation{Department of Physics, University of Duisburg-Essen, Lotharstr. 1 47057, Duisburg, Germany}

\author{Rossitza Pentcheva\orcidB}
\email{rossitza.pentcheva@uni-due.de}
\affiliation{Department of Physics, University of Duisburg-Essen, Lotharstr. 1 47057, Duisburg, Germany}

\begin{abstract}
The discovery of altermagnetism has initiated intensive research and opened new avenues for spintronic and transport applications. While current efforts are mostly focused on bulk materials which are typically insulating, here we  
%owing to its unconventional magnetic properties. Studies so far have highlighted fascinating aspects of altermagnetism, but have predominantly focused on bulk and two-dimensional materials. Also, in many cases,
%the host materials are insulators, which limits transport applications associated to  their altermagnetism. In this regard, surfaces and interfaces are particularly appealing, as they introduce additional effects associated with symmetry breaking and electronic reconstruction, yet remain largely unexplored. In this work, we demonstrate that electronic reconstruction, proximity effects of orbital ordering and oxgen octahderal rotations can stabilize interfacial metallic altermagnetism in chromate perovskite superlattices. } %Based on density functional theory calculations, we 
propose design strategies to achieve  a combination of  altermagnetism and metallicity in oxide superlattices by exploiting symmetry breaking, electrostatic doping and confinement.  %electronic reconstructionnon-relativistic spin splitting (NRSS) exploiting in perovskite chromate oxides through heterostructuring. 
While bulk SrCrO$_3$ does not exhibit altermagnetism due to compensating effects between adjacent layers our density functional theory calculations with a Hubbard $U$ parameter reveal, that a single SrCrO$_3$ layer confined in a (SrCrO$_3$)$_1$/(SrTiO$_3$)$_1$(001) superlattice (SL) exhibits a sizable non-relativistic spin splitting (NRSS)  %leading to NRSS 
up to 350 meV with bulk $d$-wave nature due to the coexistence of orbital ordering and octahedral rotations (OORs). Since this system is insulating, we extend to SrCrO$_3$/LaCrO$_3$(001) SLs.  In the (SrCrO$_3$)$_4$/(LaCrO$_3$)$_4$(001) SL the combination of a polar discontinuity at the interface and stronger OORs promotes metallic $d$-wave altermagnetism. The NRSS of up to 120 meV is contributed by the  interfacial Cr $d$ bands at the Fermi level with indications for a spin-selective Fermi surface nesting. These findings establish oxide superlattices as a promising platform to realize and explore altermagnetism  
%, but also to engineer spin-split Fermi surface suitable 
for   quantum transport and spintronic functionalities.
%\textcolor{blue}{Moreover, owing to its intrinsically two-dimensional nature, this interfacial altermagnetic state offers new opportunities for quantum transport applications and designing nanoscale devices. }

\end{abstract}

\date{\today}

\maketitle

The phenomenon of spin-polarized band splitting in compensated antiferromagnets, widely known as altermagnetism, has recently attracted considerable attention \cite{Hayami2019, Yuan2020, Yuan2021, Smejkal2022PRX, YuanZunger2023, Guo2023, Zeng2024, Lee2024PRL, Krempask2024, Reimers2024, Aoyama2024, Lin2024, Kyo-Hoon2019, Libor2020, SmejkalJairo2022, Paul2024, Bai2024, Hayami2020PRB, Jungwirth2024, Cheong2024, Radaelli2025, Bhowal2025}. Altermagnetism occurs in antiferromagnets in the presence of specific  symmetries and does not require spin–orbit coupling (SOC); the resulting spin splitting is referred to as non-relativistic spin splitting (NRSS). %due to its non relativistic origin. 
Owing to the combination of spin-splitting features—typically associated with ferromagnets—and the zero-net-magnetization character of antiferromagnets, altermagnets exhibit unconventional physical properties. These properties are particularly promising for spin transport ~\cite{Naka2019, Hernandez2021, Shao2021, Bose2022, Bai2022, Karube2021, Hu2024}, giant magnetoresistance~\cite{Libor2022}, unconventional superconductivity ~\cite{Mazin2022, Zhu2023, Banerjee2024, Chakraborty2024, Zhang2024, Lee2024}, and the emergence of chiral magnons ~\cite{Libor2023, McClarty2024, Liu2024, Morano2024, Bandyopadhyay2024}.
\\
 Perovskite oxides are particularly attractive in this context since their tunable structural distortions can also control the spin splitting. Specifically, phonon modes producing antiferro-like structural distortions can induce NRSS when coupled to suitable AFM order, as recently shown for LaMnO$_3$ and other materials~\cite{Bandyopadhyay_LMO_2025, Subhadeep_MnF2, Subhadeep_BaCuF4, Subhadeep_CuF2}. Moreover, coupling of the phonon-modes through proper structural engineering even can enable electric-field control of the NRSS, as proposed for improper ferroelectric perovskite superlattices~\cite{Bandyopadhyay_LMO_2025}. %Not only for Perovksites, this phonon-distortion mode assisted formalism works well for other antiferromagnetic materials such as rutile structured AF$_2$ (A=Mn, Cu) and BaCuF$_4$. Building of this insight, in this paper we illustrate engineering strategies to induce and control NRSS in chromate Perovskites. We shall focus on SrCrO$_3$ and LaCrO$_3$ Perovskites, which exhibit markedly different structural, electronic, and magnetic properties.
 Not only structural order, spontaneous orbital ordering (OO) can also generate NRSS ~\cite{OO_NRSS_2024}. In practice, however, OO is typically accompanied by Jahn–Teller (JT) distortions, as observed in manganite ~\cite{LMnO_OO, TMO_subhadeep}, chromate~\cite{Carta_2022}, and vanadate~\cite{AVO} perovskites. SrCrO$_3$ is one such material, that exhibits JT distortions due to the $d^2$ electronic configuration of the Cr$^{4+}$ ion. Specifically, an anti-phase $R_3^-$ JT distortion emerges  under epitaxial tensile strain  \cite{Carta_2022,AMO2} for the ground state $C$-AFM order~\cite{Chamberland_1967,SCO_Ortega_2007}, resulting in 
  an insulating phase with $G$-type OO~\cite{Carta_2022,bertino_2021}. However such a combination of AFM order and structural distortion precludes NRSS in bulk SrCrO$_3$ \cite{Bandyopadhyay_LMO_2025}, as  the contributions from adjacent layers compensate the resultant NRSS. %agreeing well with the recent report \cite{Meier_2026} that shows although NRSS is present in individual CrO$_2$ layers, 
  Meier et al.~\cite{Meier_2026} proposed to overcome this limitation by introducing additional rocksalt SrO layes  in the Ruddlesden–Popper (RP) Sr$_{n+1}$Cr$_n$O$_{3n+1}$ series. However, in these compounds, NRSS only occurs for odd values of $n$, and remains compensated in systems with even $n$.
  \\
Here we explore alternative strategies to achieve NRSS in SrCrO$_3$ through superlattice (SL) engineering. %This strategy exploits the fact that an isolated SrCrO$_3$ layer can host a net NRSS owing to the absence of an adjacent stacking layer that would otherwise compensate the net NRSS. 
We exploit the fact that an isolated SrCrO$_3$ layer can host a net NRSS by eliminating the adjacent layer that would otherwise compensate it.
  We realize this possibility in a single layer SL with the band insulator SrTiO$_3$, (SrCrO$_3$)$_1$/(SrTiO$_3$)$_1$(001), which 
  has been recently studied in a different context\cite{Rabe_SCO_STO_2015, Verma_SCO_STO_2019}. 
  Another strategy to induce  NRSS is through  engineering antiferrodistortive (AFD) oxygen octahedral rotations (OOR), which do not appear in  bulk SrCrO$_3$, due to its large tolerance factor (1.03) \footnote{The calculated tolerance factor for SrCrO$_3$ is $\approx$ 1.03 using ionic radii 1.44~\AA, 0.55~\AA\ and 1.40~\AA\ for Sr$^{2+}$, Cr$^{4+}$ and O$^{2-}$ \cite{Shannon,database_ICL} respectively.}\cite{Shannon,database_ICL}.
However, combining SrCrO$_3$ with a material hosting AFD rotations in a heterostructure may provide a promising strategy for promoting the propagation of AFD rotation patterns through the proximity effect \cite{APL_materials_2024}. LaCrO$_3$ seems an ideal candidate in this context, since in its ground-state $Pbnm$ structure \cite{OIKAWA_LaCrO3_2000,Zhou_LaCrO3_2011,LaCrO3_Tiwari} it exhibits sizable out-of-plane in-phase $M_2^+$ and in-plane anti-phase $R_5^-$ OOR (see Fig.S4 \cite{SM}). %We note that, LaCrO$_3$ itself is expected to host NRSS due to its $a^-a^-c^+$ octahedral tilt pattern\cite{Glazer} in its  $G$-AFM ground state magnetic order ~\cite{Bandyopadhyay_LMO_2025}.
\\
Beyond engineering NRSS, SLs provide an ideal platform for tailoring electronic properties through interface design and quantum confinement~\cite{Romero_2011,Lorenz_2016,Geisler_2022}. While our results show that the SrCrO$_3$-SrTiO$_3$ SLs are insulating \cite{Rabe_SCO_STO_2015, Verma_SCO_STO_2019, Tyagi_2023}, transport-based applications require metallic altermagnets with spin-split Fermi surfaces. Only a few altermagnetic metals are currently known~\cite{SmejkalJairo2022}. In this regard, thicker (SrCrO$_3$)/(LaCrO$_3$)(001) SLs offer a promising alternative, where interfacial metallic states are expected to emerge through electronic reconstruction driven by % charge transfer due to 
the polar discontinuity between the nominally charge-neutral Sr$^{2+}$O$^{2-}$/Cr$^{4+}$O$_2^{2-}$ and the polar La$^{3+}$O$^{2-}$/Cr$^{3+}$O$_2^{2-}$ planes. Our results show that the interplay of AFM order, structural distortions, and orbital degrees of freedom produces a sizable NRSS in the interfacial metallic bands, yielding a spin-split Fermi surface. The resulting quantum-confined spin-split interfacial states establish (SrCrO$_3$)/(LaCrO$_3$)(001) SLs as a promising platform for quantum transport and spintronic applications.
% where interfacial metallic states emerge through electronic reconstruction driven by charge transfer across the polar discontinuity between Sr$^{2+}$O$^{2-}$/Cr$^{4+}$O$_2^{2-}$ and La$^{3+}$O$^{2-}$/Cr$^{3+}$O$_2^{-}$ planes. On other hand, interplay of antiferromagnetic order, structural distortions, and orbital degrees of freedom produces a sizable NRSS in the interfacial metallic bands, yielding a spin-split Fermi surface in thicker (SrCrO$_3$)/(LaCrO$_3$) SL. %We demonstrate this mechanism in the (SrCrO$_3$)$_4$/(LaCrO$_3$)$_4$ (001) SL. 
 %The quantum-confined spin-split interfacial states establish these SLs as a promising platform for  quantum transport and spintronic functionalities.
 %In this context electronic reconstruction in thicker SrCrO$_3$/LaCrO$_3$ SLs. This reconstruction stems from the inter layer charge transfer to minimize the polar discontinuity  at the interface of  nominally charge-neutral Sr$^{2+}$O$^{2-}$ and Cr$^{4+}$O$_2^{2-}$ planes, and polar La$^{3+}$O$^{2-}$ and Cr$^{3+}$O$_2^{2-}$ planes, as commonly observed in perovskite SLs. 
\\
 Our density functional theory (DFT) calculations with a Hubbard correction (DFT+$U$; see Supplemental Material \cite{SM} for more details)  show that in the absence of JT distortions, SrCrO$_3$ adopts a tetragonal $P4/mmm$ structure with a ground-state $C$-AFM order and a magnetic moment of 1.95 $\mu_{\rm B}$/Cr. The $c/a$ ratio of the optimized $P4/mmm$ structure is  0.96, agreeing well with previous studies~\cite{Rabe_SCO_STO_2015}.  %Due to tetragonal distortion,  three fold degeneracy of the $t_{2g}$ orbitals is lifted, making $d_{xy}$ energetically lowered than the doubly degenerated  $d_{xz}$,$d_{yz}$ states. Presence of partially filled $d_{xz}$ and $d_{yz}$ states at the Fermi energy (E$_f$) make SrCrO$_3$ metallic. 
 However, NRSS is absent (Fig. S1\cite{SM}) %in its metallic bands 
 due to absence of $M$-point phonon distortion modes ~\cite{Bandyopadhyay_LMO_2025}.
 %, consistent with  the Ref.\cite{Bandyopadhyay_LMO_2025}. 
\begin{figure}[h]
    \centering
\includegraphics[width=\columnwidth]{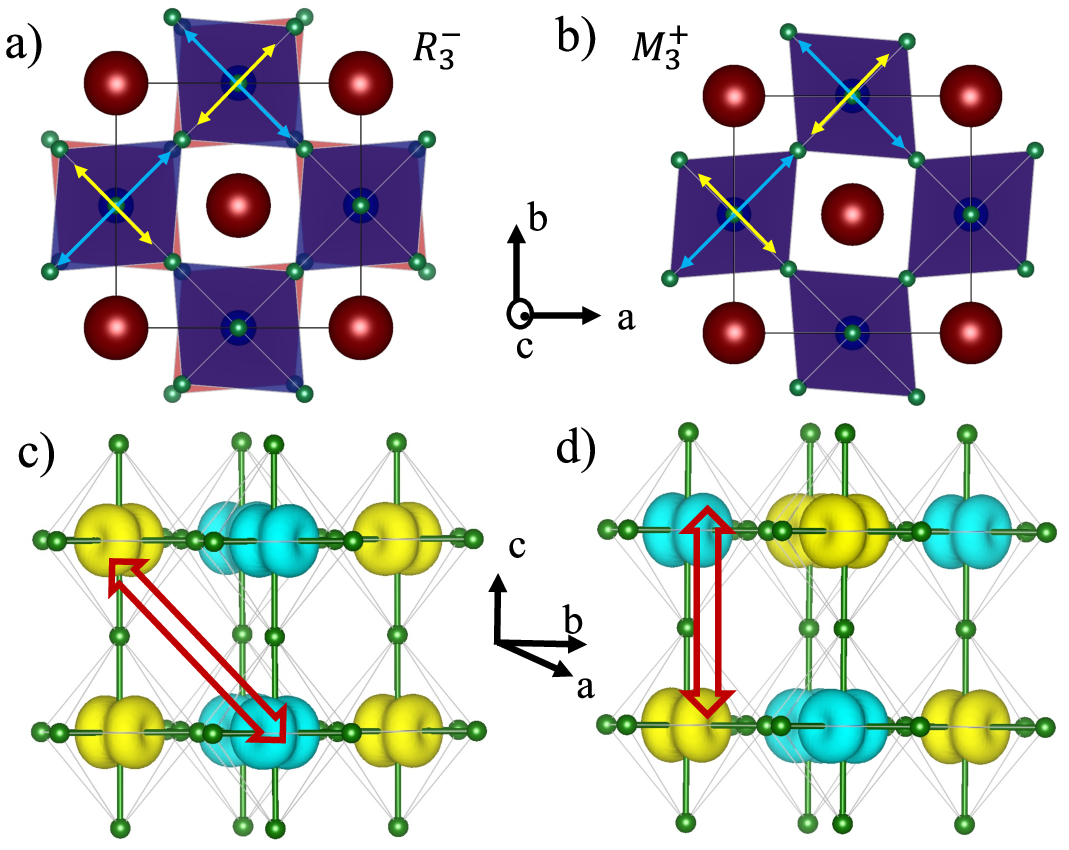}\\
    \caption{Schematic representation of a) $R_3^-$ and b) $M_3^+$ JT distortions. Blue and yellow bonds within the $ab$ plane indicate long and short bonds associated with the JT distortions. Brown, blue, green spheres  indicate Sr, Cr and O ions. Positive/negative (yellow/cyan) spin density at the Cr-sites,  for the c) $C_{\rm AF}$-$I4/mcm$ and d) $G_{\rm AF}$-$P4/mbm$ phases. A red arrow indicates the translation vector $\mathcal{T}$ connecting the oppositely aligned spin density isosurfaces.}
    \label{bulk_SCO}
\end{figure}
%However thanks to the JT distortions, that originates due to the 
\\
Introduction of JT distortions in SrCrO$_3$, which are prominent under epitaxial strain, lowers the energy, in accordance with previous studies  ~\cite{bertino_2021,Carta_2022}. %In fact, either the anti-phase $R_3^-$ or the in-phase $M_3^+$ JT distortion occurs (see Fig.~\ref{}), depending on the AFM ordering of the Cr spins ~\cite{Carta_2022}. 
Here we consider the anti-phase JT distortion $R_3^-$ and the in-phase JT distortion $M_3^+$, which are initialized through suitable oxygen displacements (see Fig.~\ref{bulk_SCO}a and b). Subsequently a structural optimization is performed both in the presence and absence of epitaxial strain for different collinear magnetic orders. 
Specifically, an in-plane tensile strain of 1.6\% is applied  to mimic the epitaxial growth conditions of SrCrO$_3$ on a SrTiO$_3$ substrate. %\footnote{For this, we constrained in-plane lattice parameters ($a$ and $b$) of SrCrO$_3$ to match those of SrTiO$_3$.}.
%Structural distortions, particularly JT distortions, may provide a viable route to induce the NRSS in SrCrO$_3$, as proposed in previous studies \cite{Bandyopadhyay_LMO_2025, Meier_2026}. In particular, the JT-active Cr$^{4+}$ ions naturally favor the development of a JT distortion. However no clear experimental evidence of a cooperative JT distortion %and the associated orbital-order–driven insulating state 
%has been reported for bulk SrCrO$_3$, but realized  in presence of epitaxial strain ~\cite{bertino_2021,Carta_2022}. 
%In this section, we investigate the $R_3^-$ and $M_3^+$ Jahn--Teller modes (see Fig.~\ref{}), initialized through suitable oxygen displacements and followed by full structural relaxation under both unstrained and epitaxially strained conditions in SrCrO$_3$. Particularly, we consider  a 1.6\%  tensile strain to mimic the epitaxial growth conditions of SrCrO$_3$ on SrTiO$_3$ substrates \footnote{For this, we constrained in-plane lattice parameters ($a$ and $b$) of SrCrO$_3$ to match those of SrTiO$_3$.}. 
We find that $C$-AFM order is preferred in the $R_3^-$ JT distorted $I4/mcm$ structure, %($Q_{R_3^-}=0.058$~\AA),
whereas $G$-AFM order is stabilized in the $M_3^+$-distorted $P4/mbm$ structure %($Q_{M_3^+}=0.082$~\AA) 
both in the absence or presence of strain, in agreement with previous reports~\cite{Carta_2022}. %The $I4/mcm$ phase  with 
The $C$-AFM ordered $I4/mcm$ phase (referred to as $C_{\rm AF}$--$I4/mcm$) is  energetically more favorable than the $G$-AFM ordered $P4/mbm$ phase (referred to as $G_{\rm AF}$-$P4/mbm$) by 1.3 meV/f.u.. Application of strain further enhances this energy difference to 14.2 meV/f.u..
%however we note that the relative stability of these JT-distorted phases depends on the considered $U$ values \cite{Carta_2022}. 
The site and orbitally projected density of states (PDOS) of the $C_{\rm AF}$-$I4/mcm$ and $G_{\rm AF}$-$P4/mbm$ phases (see Fig.~S2 ~\cite{SM})
indicates that the $d_{xy}$ orbital is completely occupied, whereas, the occupation of the $d_{xz}$ and $d_{yz}$ orbitals alternates on neighboring Cr sites. As a result, OO occurs which makes the systems insulating with band gaps of 170~meV ($C_{\rm AF}$-$I4/mcm$) and 350~meV  ($G_{\rm AF}$-$P4/mbm$). The occupied  $d_{xy}^1$$d_{xz}^1$/ $d_{xy}^1$$d_{yz}^1$ orbital configuration produces dumbbell shaped isosurfaces of spin density with different orientation at the Cr-sites, as shown in Fig.~\ref{bulk_SCO}c and d. The arrangement of the spin isosurfaces infers $G$- ($C$)-type OO in the $C_{\rm AF}$-$I4/mcm$ ($G_{\rm AF}$-$P4/mbm$) phase. %\footnote{For $C_{AF}$--$I4/mcm$ phase, this orbital arrangement alternates in both in-plane and out-of-plane, corresponding to a $G$-type OO. In contrast, the G--$P4/mbm$ phase exhibits alternating orbital occupation only within the plane, while the out-of-plane direction shows ferro-orbital alignment, corresponding to a $C$-type OO pattern.}. %Despite having JT distortion, these phases do not exhibit  NRSS in the spin polarized band structure (shown in Fig.~\ref{}), and preserve degeneracy of the spin up and down bands along all $k$-directions. 
The spin densities further reveal the presence of translational symmetry ($\mathcal{T}$) between the oppositely aligned spin sublattices for both $C_{\rm AF}$-$I4/mcm$ and  $G_{\rm AF}$-$P4/mbm$,  which precludes NRSS. %in bulk SrCrO$_3$
%the spin-polarised band structure 
 We find that NRSS appears only for the energetically unfavorable AFM order, hindering the realization of altermagnetism in bulk SrCrO$_3$ (see Fig.~S3~\cite{SM}).
 %In contrast, enforcing alternative AFM order $i.e.$ $G$- ($C$-) AFM for the $I4/mcm$ ($P4/mbm$)
  %structures enables NRSS as translation symmetry $\mathcal{T}$ breaks between the orbitally ordered oppositely aligned spin densities (see Appendix A); however, these configurations are energetically unfavorable. %\footnote{The  $G$-AFM configuration is 55 meV/f.u. higher  in energy than the and $C$-AFM configuration in  the  $I4/mcm$ structure, whereas  the  $C$-AFM configuration is 38 meV/f.u. less favorable than the $G$-AFM configuration in the  $P4/mbm$ structure }.
 %that enforce  the combination of $G$-type ($C$-type) OO with $G$-AFM ($C$-AFM) order, break translational symmetry relations and exhibit NRSS; however, these configurations are energetically unfavorable \footnote{Total energy of the  $G$-AFM configuration is 55 meV/f.u. higher than the $C$-AFM configuration for $I4/mcm$ structure. Whereas, total energy of the  $C$-AFM configuration is 38 meV/f.u. higher than the $G$-AFM configuration for $P4/mbm$ structure }.
\\
In order to overcome this limitation, we design a superlattice containing single layers of SCO and STO, %A distinct situation is realized in 
(SrCrO$_3$)$_1$/(SrTiO$_3$)$_1$(001). The ~1.6\% in-plane tensile strain on SrCrO$_3$ due to the SrTiO$_3$ layer gives rise to competing $P4/mmm$, $P4/mbm$, and $P2_1/c$   phases and FM, $C$-type AFM, and $X$-type AFM order (see Fig.~S5~\cite{SM}), consistent with previous reports
\cite{Rabe_SCO_STO_2015,Verma_SCO_STO_2019}. %Here, SrCrO$_3$ experiences a fixed in-plane tensile strain of ~1.6\% from SrTiO$_3$, that makes $P4/mmm$, $P4/mbm$, and $P2_1/c$ relevant structural phases with competing  FM, $C$-type AFM, and $X$-type AFM orders \footnote{We use a $\sqrt2\times\sqrt2\times2$ supercell containing 20 atoms to model FM and $C$-AFM order and a $2\times2\times2$ supercell containing 40 atoms to model $X$-AFM for all considered structures.} (shown in Fig.\ref{}).
The $P4/mmm$ structure lies higher in energy than  the $P4/mbm$ and $P2_1/c$ phases, independent of the magnetic order considered. Although $P4/mbm$ and $P2_1/c$ are nearly degenerate, the latter is lower in energy by $0.8$ meV/f.u.\ in its preferred $C$-AFM configuration.
%$P4/mmm$ structure does not incorporate JT distortion, whereas both $P4/mbm$ and $P2_1/c$ incorporate the
The $M_3^+$ JT distortion, common to both $P4/mbm$ and $P2_1/c$ structures, promotes OO in the SrCrO$_3$ layer. In the presence of $C$-AFM order this OO breaks   $\mathcal{T}$ symmetry between the antiparallel spin densities. Consequently, a $d$-wave NRSS up to 350 meV emerges along the $k_xk_y$ direction (see Fig.~\ref{SCO_STO}a and Fig.~S6~\cite{SM}).
 %The $M_3^+$ JT distortion promotes OO in the SrCrO$_3$ layer for both the $P4/mbm$ and $P2_1/c$ structures. For the  ground state $C$-AFM order, the resulting spin densities break $\mathcal{T}$ symmetry between the antiparallel spin sites, leading to  a sizable NRSS along the $k_xk_y$ direction (see Fig.~\ref{SCO_STO}a and Appendix B). 
 The $P2_1/c$ phase additionally possesses small antiphase OORs (2.1$^\circ$ along $x$ and $y$ ($\phi^-_{xy}$) and 0.6$^\circ$ along the $z$ ($\phi^-_z$) axis), inducing a $a^-a^-c^-$ octahedral tilt pattern.  This leads to an additional $d$-wave NRSS along  $k_xk_z$ which is  weaker compared to the $k_xk_y$ direction (Fig.~S6~\cite{SM}). Importantly, NRSS along two distinct $k$-directions, makes the $P2_1/c$ phase altermagnetic with bulk $d$-wave nature \cite{Smejkal2022PRX,Subhadeep_CuF2}.
 %However, these phases remain insulating preserving the electronic properties of the bulk parent compounds
 This highlights the role of confinement in the SL together with   OOR  arising from the octahedral connectivity through the interface to  induce a robust altermagnetism, which is quenched in the bulk parent compound SCO. %which can be  generated in SrCrO$_3$ through SL engineering. \\%In (SrCrO$_3$)$_1$/(SrTiO$_3$)$_1$ SL, however the ORs are very small and the $M_2^+$ mode is even absent. %Although being a $M$-point phonon mode, $M_2^+$ is expected to play a crucial role in emerging the NRSS for the $C$-AFM order. 
\begin{figure}[h!]
\centering
\includegraphics[width=\columnwidth]{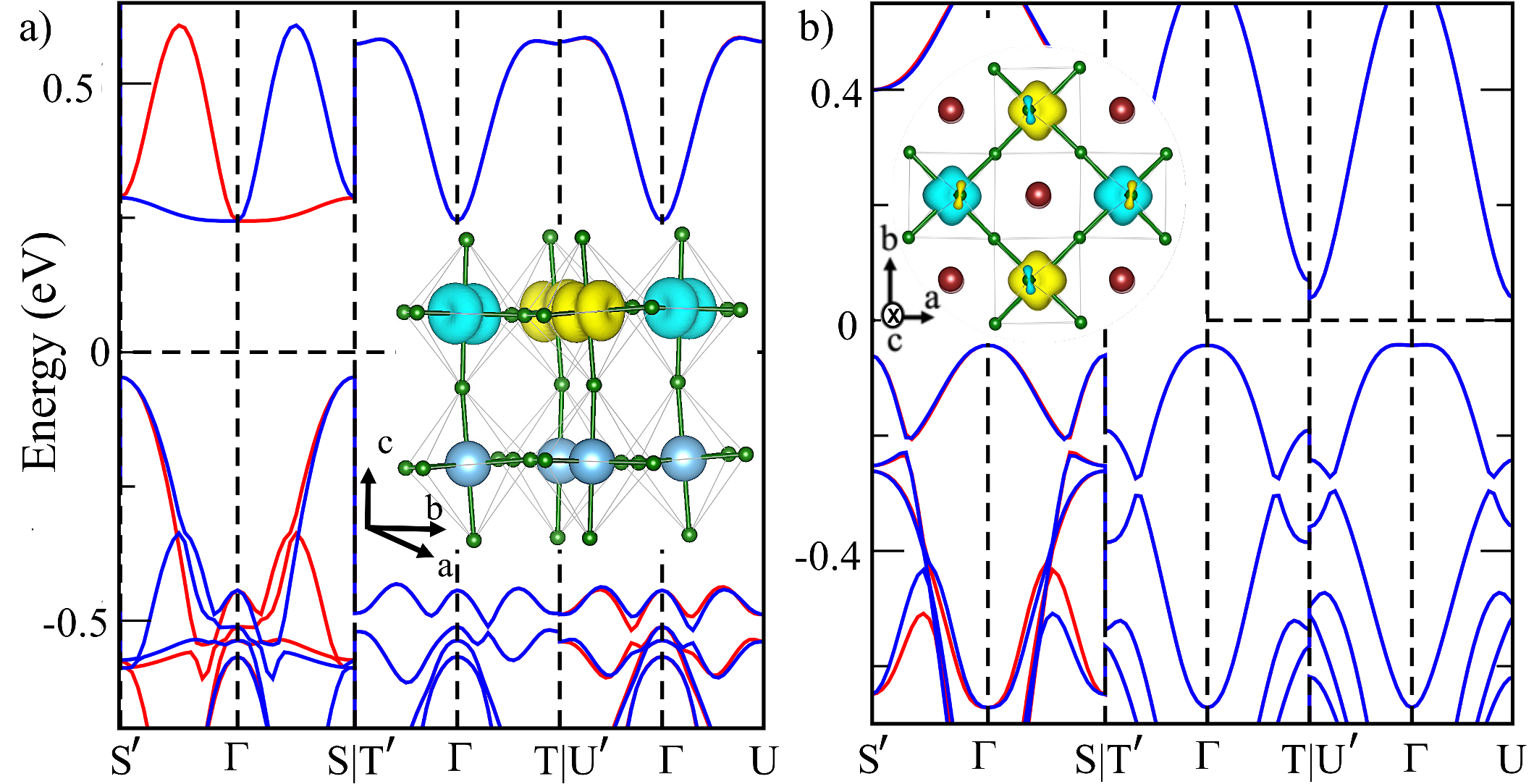}\\
\caption{Spin polarised band structure of  a)  $(\mathrm{SrCrO_3})_1/(\mathrm{SrTiO_3})_1$(001) and b) $(\mathrm{SrCrO_3})_1/(\mathrm{LaCrO_3})_1$(001) SLs. Red and blue lines represent spin up and down bands, high-symmetry k-points are given by $\rm{S^{'}/S}$~($\mp$0.5,0.5,0), $\Gamma$(0,0,0), $\rm{T^{'}/T}$~(0,$\mp$0.5,0.5) and $\rm{U^{'}/U}$~(0.5,0,$\mp$0.5). The insets show the corresponding spin density of the Cr sites (yellow/cyan for up/down) in their resp. crystal structures.}
\label{SCO_STO}
\end{figure}
%Since the SrTiO$_3$/SrCrO$_3$(001) SL remains insulating  and the OORs are small and the $M_2^+$ mode  even absent, we now turn to LaCrO$_3$/SrCrO$_3$(001) SLs with the aim of inducing stronger OORs into the SrCrO$_3$ (SCO) layer in proximity of LaCrO$_3$ (LCO).
\\
To induce stronger OORs, we now turn to SrCrO$_3$/LaCrO$_3$ SLs.
Beyond their structural differences, SrCrO$_3$ and LaCrO$_3$ differ markedly in their electronic and magnetic properties. Our results show that unlike SrCrO$_3$, bulk LaCrO$_3$ with its $G$-type AFM ground state and $Pbnm$ structure~\cite{OIKAWA_LaCrO3_2000,Zhou_LaCrO3_2011,LaCrO3_Tiwari} hosts NRSS along the $k_xk_z$ direction (see Fig.~4~\cite{SM}), owing to the presence of an $a^-a^-c^+$ octahedral tilt pattern~\cite{Bandyopadhyay_LMO_2025}. Second, the insulating nature of LaCrO$_3$ arises from the fully occupied majority-spin  $t_{2g}$ states
%yielding a magnetic moment of 2.76~$\mu_{\rm B}$/Cr, consistent with of the nominal Cr$^{3+}$ valence,
  of the Cr$^{3+}$ ions,
 while rendering the JT effect negligible. Besides these electronic differences of the bulk constituents, a central feature is the valence discontinuity at the interface of the SrCrO$_3$/LaCrO$_3$ SLs
which leads to a charge transfer and  exhibits a strong thickness dependence. % of the SLs~\cite{}.  
To get further insight, we systematically explore  the evolution of the structural, electronic and altermagnetic properties in (SrCrO$_3$)$_n$/(LaCrO$_3$)$_n$(001) SLs  with $n=1,~2,~4$ layers of SrCrO$_3$ and LaCrO$_3$, denoted as ($n$,$n$). 
\\
\begin{figure}[t!]
    \centering
\includegraphics[width=\columnwidth]{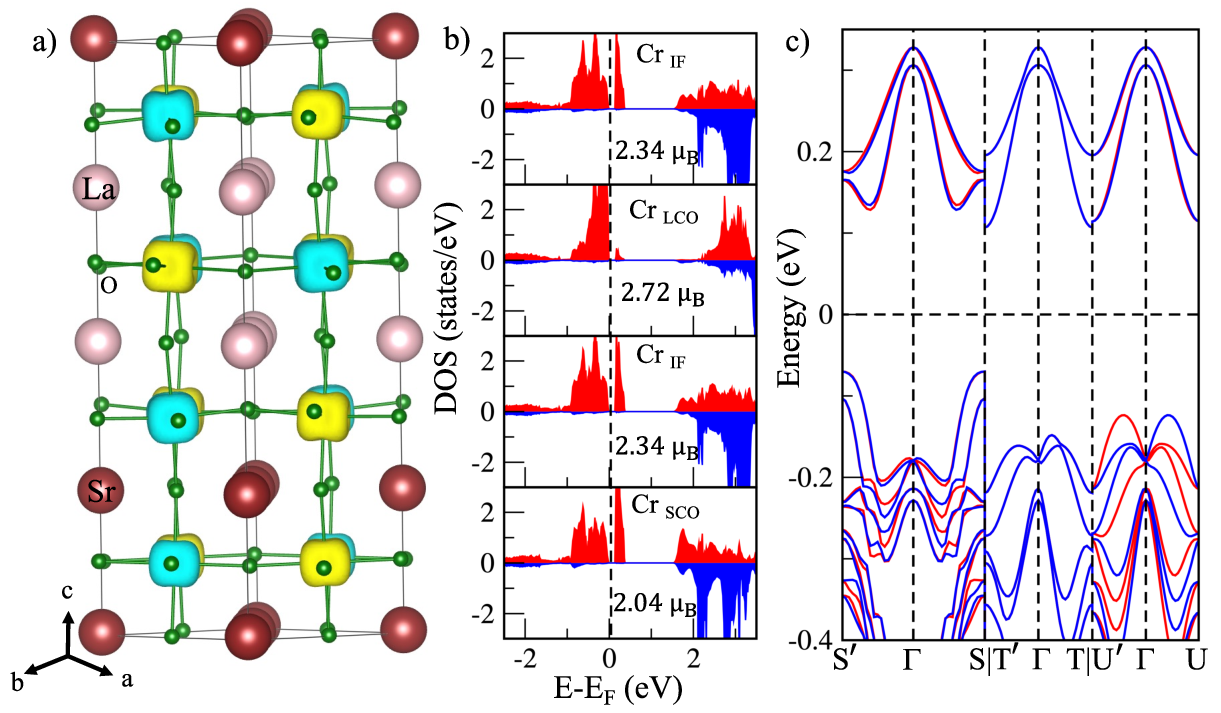}\\
    \caption{(a) Crystal structure of the  (SrCrO$_3$)$_2$/(LaCrO$_3$)$_2$ (001) SL showing the spin-density distribution for the $CG$-AFM ground state. Brown, beige, and green spheres represent Sr, La, and O ions, respectively, while yellow and cyan isosurfaces (isovalue 0.035 e/\AA$^3$) denote spin-up and spin-down densities. (b) PDOS of the Cr-$d$ states for Cr sites in different layers. Red (blue) color correspond to spin up (down) states. (c) Spin-polarized band structure. Red/blue lines indicate spin-up/down bands. }
    \label{SCO_LCO_2_2}
\end{figure}

\begin{figure*}[t!]
    \centering
\includegraphics[scale=0.27]{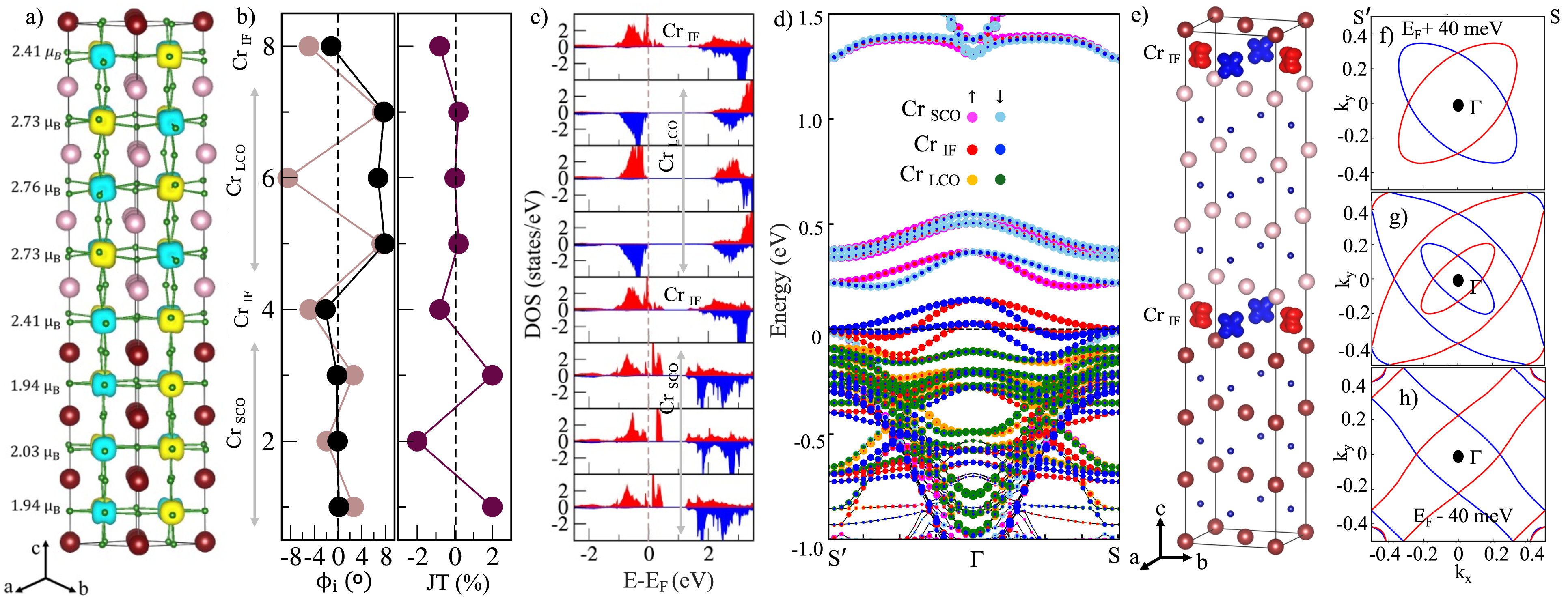}\\
    \caption{(a) Crystal structure of the  (SrCrO$_3$)$_4$/(LaCrO$_3$)$_4$ (001) SL showing the spin-density distribution for the $CG$-AFM ground state. Brown, beige, and green spheres represent Sr, La, and O atoms, respectively, while yellow and cyan isosurfaces (isovalue = 0.035 e/\AA$^3$) denote spin-up and spin-down densities. b) Layer-resolved variation of the OOR and JT distortion amplitudes at the Cr sites. Beige (black) circles represent in-plane (out-of-plane) OOR amplitudes, while maroon circles denote the JT distortion amplitudes. (c) PDOS of the Cr-$d$ states for Cr sites in different layers. Red (blue) curves correspond to spin up (down) states. (d) Spin-polarized band structure  along the $\rm S^{'}(-0.5,0.5,0)-\Gamma(0,0,0)-\rm S(0.5,0.5,0)$ path. Red/blue lines indicate spin-up/down) bands. Cr-$d$ orbitals are projected on the bandstructure, where, magenta, red, and orange (Light blue, blue, and green) symbols represent contributions from spin up (down) Cr-$d$ states associated with the SrCrO$_3$ layers (layers 1–3), interfacial layers (layers 4 and 8), and LaCrO$_3$ layers (layers 5–7), respectively. e) Band decomposed  spin density (red/blue isosurfaces are for up/down spin) for the bands crossing E$_{\rm F}$. Spin split Fermi surfaces in $k_x$-$k_y$ plane ($k_z$=0) at f) E$=$E$_{\rm F}+$40 meV, g) E=E$_{\rm F}$ and h) E=E$_{\rm F}-$40 meV.}
    \label{fig_44_SL}
\end{figure*}
%We consider OORs, $M_3^+$ and $R_3^-$ JT distortions during the structural optimization for different magnetic orders  to identify the most favorable phase \cite{SM}.
%$\mathrm{(SCO)}_n/(\mathrm{LCO})_n$(001) SL structures were initialized with an overall $Pbnm$-like OORs pattern, then $M_3^+$ and $R_3^-$ JT distortion modes were introduced. Finally, a structural optimization was performed for different magnetic orders to identify the most favorable phase. % (see SM for details).
 In the $(\mathrm{SrCrO_3})_1/(\mathrm{LaCrO_3})_1$(001) SL $Pbnm$-like OORs lower the symmetry to orthorhombic $Pmc2_1$, where $M_3^+$ JT distortion is symmetry allowed and can occur naturally. On the other hand, inclusion of the $R_3^-$ JT mode  lowers the symmetry further to monoclinic $Pc$. We note that  the lattice mismatch between LaCrO$_3$ and SrCrO$_3$ generates a 
 $\approx$0.5\% tensile strain on SrCrO$_3$, which favors JT distortions in the SL.
 %Structural optimization of the 
The $Pmc2_1$ structure is stabilized in $(\mathrm{SrCrO_3})_1/(\mathrm{LaCrO_3})_1$(001) for FM, $A$-AFM and $C$-AFM order, whereas $G$-AFM stabilizes a $Pc$ structure. We find that the  $Pmc2_1$ structure with  $C$-AFM order (referred to as $C_{\rm AF}$-$Pmc2_1$ phase) is most favorable \footnote{ 
 The optimized $Pmc2_1$ structure with $A$-AFM, and FM order lie 45, and 71 meV/f.u. higher in energy than the $C_{\rm AF}$-$Pmc2_1$ ground state, respectively.}, while the $Pc$ structure with $G$-AFM order (referred to as $G_{\rm AF}$-$Pc$ phase) is 25 meV/f.u. higher in energy. %An overall $\approx$0.5\% tensile strain applies on SrCrO$_3$ due to the lattice mismatch with LaCrO$_3$, which generates JT distortions in $(\mathrm{SrCrO_3})/(\mathrm{LaCrO_3})$ SLs. 
 %Unlike the $(\mathrm{SrCrO_3})_1/(\mathrm{SrTiO_3})_1$ $P2_1/c$ structure,
  The $C_{\rm AF}$-$Pmc2_1$ phase gives rise to a  $a^-a^-c^+$ octahedral tilt pattern, induced by the $R_5^-$ and $M_2^+$ OOR modes. Additionally, $C_{\rm AF}$-$Pmc2_1$ hosts the $M_3^+$ JT and some other distortion modes (see SM~\cite{SM}). Among them,  $X_1^+$ comprising the antipolar displacement of the Cr atoms along  $z$ is significant, whereas the $M_2^+$ OOR ($\phi_z^+=0.8^\circ$) and $M_3^+$ JT modes are weak.
 The four equivalent Cr sites in the \(C_{\rm AF}\)-\(Pmc2_1\) phase carry magnetic moments of \(2.4~\mu_{\rm B}\)/Cr,  indicating an intermediate Cr\(^{3.5+}\) valence arising from charge transfer between  LaCrO$_3$ and SrCrO$_3$. %layers driven by the valence discontinuity. 
 The resulting PDOS shows an insulating state with  a band gap of 0.08 eV between occupied and unoccupied  states with mixed Cr-$d_{yz}$ and $d_{xz}$ character, induced by the joint action of the $R_5^-$ and $X_1^+$ modes ~\cite{SM}.
Importantly, due to the $a^-a^-c^+$ octahedral tilt pattern, the $\mathcal{T}$ symmetry is broken between the AFM sublattices %with similar orbital occupancies 
 (see spin densitites in Fig.~\ref{SCO_STO}b). This gives rise to planar $d$-wave altermagnetism with NRSS up to 80 meV along the $k_xk_y$ direction, %for the valence bands 0.5 eV below $E_{\rm F}$, 
whereas the  NRSS along  $k_xk_z$ of bulk LaCrO$_3$ is absent here. %The energy difference between the split bands are small, likely due to the weak $M$-point phonon distortion modes: $M_2^+$ and  $M_3^+$.
%Still, the system remains an insulator, where a band gap of 0.08 eV is opened between states having mixed Cr-$d_{yz}$ and $d_{xz}$ character, which are weakly polarized due to the weak $M_3^+$ JT. %While, the Cr-$d_{xy}$ is fully occupied and $e_g$ states are empty.  
On the other hand,  the metastable $G_{\rm AF}$-$Pc$ phase turned out to be metallic, where two inequivalent Cr sites with magnetic moments of 2.6 and 2.1~$\mu_{\rm B}$, exhibit 
a layered ordering along the crystallographic $c$ direction.
%layered ordering of the associated  Cr$^{3+}$- and Cr$^{4+}$-like charge states  along the $z$ direction. 
The $G_{\rm AF}$-$Pc$ structure exhibits an $a^-a^-c^-$ octahedral tilt pattern %with $\phi_z=3.7^\circ$, 
as well as  stronger $M_3^+$, $R_3^-$ JT distortions. %The stronger JT distortion at the Cr$^{4+}$ sites stabilizes $d_{xz}/d_{yz}$ orbital order, whereas the nearly  Cr$^{3+}$ sites yield a metallic state. 
The $G$-AFM order in presence of the mentioned structural distortions enables  bulk $d$-wave altermagnetism in the $G_{\rm AF}$-$Pc$ phase with NRSS up to 220 meV along the $k_xk_y$ and up to 35 meV along the $k_xk_z$ directions, which is larger  than the NRSS of the $C_{\rm AF}$-$Pmc2_1$ phase (see SM~\cite{SM}). However being energetically unfavorable this phase may not be accessible.   
\\
Next we move to the thicker (SrCrO$_3$)$_2$/(LaCrO$_3$)$_2$(001) SL, where a combination of $C$ and $G$-AFM magnetic order of Cr in the LaCrO$_3$ and SrCrO$_3$ layers, respectively, is preferred. We refer to this fully compensated AFM  order as $CG$-AFM  
\footnote{Other  AFM order  at the Cr-sites, such as $G$-AFM, $C$-AFM for the whole SL were energetically less favorable}. The optimized (2,2) SL  shows large OORs in the LaCrO$_3$ layers that gradually decrease towards the SrCrO$_3$ layers. The magnetic moments of 2.3, 2.0 and 2.7 $\mu_{\rm B}$ indicate an intermediate ${3.5+}$ state at the interfacial Cr$_{\rm IF}$ ions, while Cr sites farther from the interface in the %central bulk-like 
SrCrO$_3$ (Cr$_{\rm SCO}$) and LaCrO$_3$  (Cr$_{\rm LCO}$) layers preserve bulk-like Cr$^{4+}$ and Cr$^{3+}$ character. %, %with giving rise to the magnetic moments of 2.3, 2.0 and 2.7 $\mu_{\rm B}$ respectively. 
The JT distortion~\footnote{The JT distortion at each Cr layer is quantified using the ratio of the in-plane O--Cr--O bond lengths.} is present at the Cr$_{\rm SCO}$ and Cr$_{\rm IF}$ sites, %significant at the Cr$_{\rm IF}^{3.5+}$  sites, but becomes  weaker  for the Cr$_{\rm SCO}^{4+}$ sites and 
%eventually gets suppressed  
but absent at the Cr$_{\rm LCO}^{3+}$ sites. Consequently,  the Cr$_{\rm IF}^{3.5+}$ sites exhibit $d_{yz}$/$d_{xz}$ OO as evident from the spin density (see Fig.~\ref{SCO_LCO_2_2}a-b), leading to an insulating state with a band gap of 130 meV. The interplay of structural, orbital and magnetic order leads to  NRSS   along both the $k_xk_y$ and $k_xk_z$ directions, as shown in Fig. \ref{SCO_LCO_2_2}c.
%The NRSS along the $k_xk_z$ is prominent in the valence bands, which are predominantly driven by the $t_{2g}$ states of Cr$_{LCO}^{3+}$ owing $G$-AFM order across the neighboring Cr sites, consistent with the feature of bulk LaCrO$_3$.While the NRSS along the $k_xk_y$ are relatively small and contributed by the  $t_{2g}$ states of Cr$_{SCO}^{4+}$ and  Cr$_{IF}^{3.5+}$ atoms having a $C$-AFM order. 
%An overall $\approx$0.5\% tensile strain applies on SrCrO$_3$ due to the lattice mismatch with LaCrO$_3$, which generates JT distortions in $(\mathrm{SrCrO_3})/(\mathrm{LaCrO_3})$ SLs. 
In practice, these SLs are typically grown on substrates, which impose an additional epitaxial strain. To investigate the effect of epitaxial strain on the electronic and magnetic properties of the SrCrO$_3$/LaCrO$_3$ SLs, we fix the in-plane lattice constant to that of SrTiO$_3$, a widely used substrate for the epitaxial growth of SrCrO$_3$ and LaCrO$_3$ thin films \cite{SCO_on_STO,LCO_on_STO}. This corresponds to a tensile strain of approximately 1.6\% for the SrCrO$_3$ layers and 0.6\% for the LaCrO$_3$ layers. Strain further stabilizes the JT distorted ground state, for example, the energy difference between the \(C_{\rm AF}\)-\(Pmc2_1\) and $G_{\rm AF}$-$Pc$ is
enhanced to 32 meV/f.u. for (SrCrO$_3$)$_1$/(LaCrO$_3$)$_1$ SL, making the former energetically more stable.  Importantly, despite structural modifications, the overall altermagnetic properties of the SLs remain largely unaffected by strain. Therefore, the key findings obtained for the unstrained structures remain valid under SrTiO$_3$-induced tensile strain.
\\
We now extend our analysis to the (SrCrO$_3$)$_4$/(LaCrO$_3$)$_4$(001) SL,  which facilitates a clearer distinction between bulk-like and interfacial regions, allowing us to better resolve their respective contributions to the NRSS. Here we consider epitaxial tensile-strain of  SrTiO$_3$ for the structural optimization. Similar to the (2,2) SL, $CG$-AFM order   is preferred also for (4,4) with $C$-AFM order of the Cr sites in the SCO layers and $G$-AFM order in the LCO part,  %The comparison of different magnetic orders further confirms stability of the $CG$-AFM order, in which the Cr sites across the SrCrO$_3$ layers (Cr$_{SCO}$) exhibit  $C$-AFM order, and the  Cr sites across the LaCrO$_3$ layers (Cr$_{LCO}$)  exhibit a $G$-AFM order, whereas Cr sites at the two interfaces (Cr$_{IF}$) experience both $C$ and $G$-AFM while interfacing Cr$_{SCO}$ and Cr$_{LCO}$ respectively
as evident from the associated spin density, shown in Fig.~\ref{fig_44_SL}a. $CG$-AFM corresponds  to  a fully compensated AFM order with net zero total magnetic moment. The magnetic moments of 1.94–1.99 $\mu_{\rm B}$/Cr$_{\rm SCO}$ and 2.73–2.76 $\mu_{\rm B}$/Cr$_{\rm LCO}$ infer bulk-like Cr$^{4+}_{\rm SCO}$ and Cr$^{3+}_{\rm LCO}$ charge states, respectively. In contrast, the magnetic moment of 2.41 $\mu_{\rm B}$/Cr$_{\rm IF}$ indicates a mixed-valence Cr$^{3.5+}$ state at the interface. Moreover, structural analysis  reveals that OOR and JT distortions are the primary structural order parameters that have strong layer-dependence as shown in  Figure \ref{fig_44_SL}b.  Alternating (same) signs of the OOR amplitude in adjacent Cr layers correspond to antiphase (in-phase) AFD OORs. The OOR amplitude is large in the LaCrO$_3$ layers and smaller in the SrCrO$_3$ layers, consistent with their bulk properties. Likewise, the JT distortion is pronounced in the SrCrO$_3$ layers but relatively weak in the LaCrO$_3$ layers. The alternating sign of the JT distortion between the adjacent SrCrO$_3$ layers indicates the appearance of the $R_3^-$ JT mode and is consistent with the development of a $G$-type OO, as evident at the  Cr$_{\rm SCO}$ sites in Fig.~\ref{fig_44_SL}a. Notably, the interfacial Cr layers display intermediate OOR and JT amplitudes, indicative of the competing influence of the neighboring environments.
%induces G-type OO at the Cr$^{4+}$ sites. SrCrO$_3$ layers exhibit magnetic moments of 1.94–1.99 $\mu_B$/Cr, while, Cr$^{3+}$ sites with magnetic moments of 2.73–2.76 $\mu_B$/Cr stabilize a G-AFM order across the LaCrO$_3$ layers, thereby retaining their bulk-like electronic character. The four Cr sites at the two interfacial layers exhibit  magnetic moments of 2.41 $\mu_B$/Cr,  Importantly, AFM order within individual Cr layers leads to a fully compensated antiferromagnetic configuration. 
Consequently, the charge transfer at the interface in conjunction with the structural and magnetic order drives the system metallic, as visible in the PDOS and band structure in Figs.~\ref{fig_44_SL}c and d. Metallicity is exclusively contributed by the interfacial Cr$_{\rm IF}$ $t_{2g}$ bands, while the fully occupied $t_{2g}$ states of Cr$^{3+}_{\rm LCO}$ and the  JT distorted $t_{2g}^2$ states of Cr$^{4+}_{\rm SCO}$ remain gaped, predominantly contributing to the valence and conduction bands. The band structure showing the layer-dependent Cr-$d$ character in Fig.~\ref{fig_44_SL}d further confirms  the dominant contribution of  the Cr$_{\rm IF}$-$d$ states around $E_{\rm F}$. Most importantly, the resulting band structure clearly shows  NRSS of $d$-wave nature along the $k_xk_y$  and $k_xk_z$ (see Fig.~ S10 \cite{SM}) directions. The  valence and conduction bands along the $k_xk_y$ direction exhibit a weak NRSS, attributed to the $G$-OO and $C$-AFM order of the Cr$^{4+}_{\rm SCO}$ and the $G$-AFM order of Cr$^{3+}_{\rm LCO}$, since both of these combinations preclude NRSS along the $k_xk_y$ direction in the bulk counterparts. In contrast, a significant NRSS of up to 120 meV emerges for the bands crossing the Fermi level, contributed predominantly by Cr$_{\rm IF}$. %, which is very significant. 
While the NRSS of the Cr$_{\rm SCO}$(Cr$_{\rm LCO}$) bands stems from the uncompensated JT distortion and OORs between the adjacent Cr layers, the NRSS of the Cr$_{\rm IF}$ at the Frmi level originates from the partial OO [see alternating $d_{xz}$/$d_{yz}$ occupation at  neighboring  Cr$_{\rm IF}$ sites in  Fig.~\ref{fig_44_SL}(e)], induced OOR and combination of $C$ and $G$-AFM order of the neighboring Cr sites. 
%Most importantly, we reveal that the metallic states are localized at the interfacial Cr sites, as confirmed by the spin density in Fig.~\ref{fig_44_SL}(e). 
The nearly dispersionless bands along the $k_z$ direction (see Fig.~ S10 \cite{SM}), further highlight the quasi two-dimensional nature of the metallic spin-split bands. Consequently, this gives rise to two pairs of elliptic spin-split Fermi surfaces (FSs) (Fig.~\ref{fig_44_SL}g), confined to the interfacial Cr layers, making this system a promising platform for quantum transport applications. We notice significant changes in the FS topology at different energy levels. A single pair of spin-split FSs emerge at approximately 40 meV above and below E$_{\rm F}$ (Fig. \ref{fig_44_SL}f,h). Notably, below E$_{\rm F}$, the cigar-shaped FS suggests possible spin selective nesting, which may promote unconventional superconductivity under applied bias.
\\
Finally, we investigate the dependence of the predicted interface-induced NRSS w.r.t variation of the Hubbard $U$ parameter.  Our results with a larger U$_{\rm eff}$= 3 eV show similar interfacial Cr-$d$  metallic spin-split bands crossing the Fermi level~\cite{SM}, proving the robustness of our proposed mechanism even for stronger corelation. 
 % whereas the valence and conduction bands are mainly derived from the Cr$^{3+}_{\rm LCO}$ and Cr$^{4+}_{\rm SCO}$ respectively. 
%Surprisingly, in contrast to the prohibiting effects of $G$-OO and $C$-AFM in bulk SrCrO$_3$, which prevent the appearance of a NRSS, and the $G$-AFM phase of bulk LaCrO$_3$, which suppresses the NRSS along the $k_xk_y$ direction, the band structure of the SL (Fig.\ref{fig_44_SL}d) shows a NRSS along the $k_xk_y$ direction.  
%The spin splitting has a $d$-wave nature, where the splitting  is found larger for the Cr$_{\rm IF}$ derived metallic bands than the Cr$_{\rm SCO}$ (Cr$_{\rm LCO}$) dominated conduction (valence) bands. We have also explored the effect of stronger electronic correlatio
\\
$Conclusions-$ In a comprehensive DFT+$U$ study we develop strategies to induce NRSS in SrCrO$_3$-derived perovskite SLs. Bulk SrCrO$_3$ intrinsically precludes altermagnetism because the ground state  combination of AFM  and orbital order is incompatible with the symmetry requirements for NRSS. In contrast, the confinement of a single SrCrO$_3$ layer in the (SrCrO$_3$)$_1$/(SrTiO$_3$)$_1$(001)  SL stabilizes a $C$-type OO, which, together with the favored $C$-type AFM order, gives rise to NRSS up to 350 meV. In SrCrO$_3$/LaCrO$_3$(001) SLs, OORs introduced by the LaCrO$_3$ layers, play a crucial role in enabling NRSS. Furthermore, (SrCrO$_3$)$_n$/(LaCrO$_3$)$_n$ (001) SLs, $n$= 2, 4, exhibit NRSS along two distinct $k$-space directions resulting in a bulk $d$-wave altermagnet. Most importantly, electronic reconstruction in the (SrCrO$_3$)$_4$/(LaCrO$_3$)$_4$(001) SL due to the polar discontinuity at the interace gives rise to metallic interfacial Cr $d$ states with sizable NRSS, leading to spin-split Fermi surfaces, a key ingredient for potential spin-transport applications of altermagnets. Our findings may be relevant for a broader class of  orbitally ordered antiferromagnetic oxides such as vanadate and manganite perovskites and establish oxide superalttices as a rich platform to search for and explore robust metallic altermagnets.
\\
$Acknowledgements-$ The authors gratefully acknowledge the use of computational resources provided by the amplitUDE supercomputer at the
University of Duisburg-Essen (DFG grant INST 20876/423-1 FUGG).

\bibliography{main}

\end{document}